
\input harvmac
\def\Title#1#2#3#4{\nopagenumbers\abstractfont\hsize=\hstitle\rightline{#1}%
\vskip 1in\centerline{\titlefont #2}
\vskip 10pt
\centerline{\titlefont #3}
\vskip 10pt
\centerline{\titlefont #4}
\abstractfont\vskip .5in\pageno=0}

\def\npb{{Nucl.\ Phys.\ }{\bf B}}
\def\physrep{Phys.\ Reports\ }
\def\plb{{Phys.\ Lett.\ }{\bf B}}

\def\prd{{Phys.\ Rev.\ }{\bf D}}
\def\prl{Phys.\ Rev.\ Lett.\ }

\def\zpc{Z.\ Phys.\ {\bf C}}

\def \b{\beta}
\def \a{\alpha}
\def \g{\gamma}
\def \d{\delta}
\def \e{\epsilon}
\def \l{\lambda}
\def \th{\theta}
\def \ph{\phi}

\def \Ph{\Phi}

\def \m{\mu}

\def\5bar{\bar 5}
\def\10bar{\bar 10}
\def\Hbar{\bar H}
\def\q{\quad}

\def\sy{supersymmetry}
\def\sic{supersymmetric}

\def\ssm{supersymmetric standard model}
\def\sm{standard model}

\def\lf{16\pi^2}
\def\llf{(16\pi^2)^2}
\Title{LTH 345}{Renormalisation-group invariance}{and}
{universal soft supersymmetry-breaking}
\centerline{I. Jack and D. R. T. Jones}
\bigskip
\centerline{\it DAMTP, University of Liverpool, Liverpool L69 3BX, U.K.}
\vskip .3in

We show that a particular ``universal'' form for the soft-breaking couplings
in a softly broken $N=1$ supersymmetric gauge theory is renormalisation-group
invariant through two loops, provided we impose one simple condition on
the dimensionless couplings. The universal form for the trilinear couplings
and mass terms is identical to that found in popular derivations of the
soft-breaking terms from strings or supergravity.
\Date{January 1995}

If we take the \sm , generalise to two Higgs doublets, supersymmetrise,
impose R-parity, and add all possible soft \sy\ breaking terms  then we
have the \ssm . The resulting theory has an alarming number of
arbitrary parameters; far more than the \sm .  It is customary to assume
that the plethora of possible  independent soft terms undergo a form of
unification,  at the same scale where the gauge couplings meet. At this
scale it is supposed  that the soft terms consist simply of a common
scalar mass, a  common gaugino mass, and $\phi^3$ and $\phi^2$
interactions proportional  to the analogous terms in the superpotential;
the constants of proportionality  being denoted $A$ and $B$
respectively. This simplification can be motivated to some extent
by appeal to $N=1$ supergravity, and in particular to the idea that
the \sy\ breaking occurs in a hidden sector and is communicated to
the observable sector via gravitational interactions
(for a review, see \ref\sug{H.-P.~Nilles, \physrep  {\bf C}110 (1984) 1.}).
It also arises
in superstring phenomenology\ref\IL{L.E. Ib\'a\~nez and D. L\"ust,
\npb382 (1992) 305\semi
V. Kaplunovsky and J. Louis, \plb306 (1993) 269\semi
R.~Barbieri, J.~Louis and M.~Moretti, \plb 312 (1993) 451;
erratum-{\it ibid\/} 316 (1993) 632.}
\ref\CM{A. Brignole, L. E. Ib\'a\~nez and C. Mu\~noz, \npb422 (1994) 125.}.

In this note we attempt to motivate a simple form for the soft breakings in
a different way. We explore the consequences of imposing that
the soft breakings in the theory at the unification scale be form
invariant under renormalisation. In other words we require  that the
theory be renormalisable, in the usual sense that counter-terms
generated by shifting parameters and fields in the
Lagrangian suffice to remove the divergences encountered in
perturbation theory.
 In general, of course, imposing
strict renormalisability requires us to write down all interactions
permitted by the symmetries. We will find, however, that
a particular universal form for the
soft-breaking couplings (one which is compatible with the desired
pattern of \sy\ breaking described above)
is renormalisation-group (RG) invariant at least through two loops
provided we impose one simple
condition on the dimensionless coupling sector of the theory.
Theories with this property would have the attractive feature that the
universal form of the soft breaking terms (which is presumably generated
by supersymmetry breaking of the underlying supergravity or superstring theory
at or near the Planck scale) would be exactly preserved down to the
gauge unification scale.
The Lagrangian $L_{\rm SUSY} (W)$ is defined by the superpotential
\eqn\Aae{
W={1\over6}Y^{ijk}\Ph_i\Ph_j\Ph_k+{1\over2}\m^{ij}\Ph_i\Ph_j+L^i\Ph_i.}
$L_{\rm SUSY}$ is the Lagrangian for  the $N=1$ supersymmetric
gauge theory, containing the gauge multiplet $\{A_{\m},\l\}$ ($\l$ being the
gaugino) and a chiral superfield $\Ph_i$ with component fields
$\{\ph_i,\psi_i\}$ transforming as a (in general reducible)
representation $R$ of the gauge group $\cal G$.
We assume that there are no gauge-singlet
fields and that $\cal G$ is simple. (The generalisation to a semi-simple
group is trivial.)
The soft breaking is incorporated in $L_{\rm SB}$, given by
\eqn\Aaf{
L_{\rm SB}=(m^2)^j_i\ph^{i}\ph_j+
\left({1\over6}h^{ijk}\ph_i\ph_j\ph_k+{1\over2}b^{ij}\ph_i\ph_j
+ {1\over2}M\l\l+{\rm h.c.}\right)}
(Here and elsewhere, quantities with superscripts are complex conjugates of
those with subscripts; thus $\ph^i\equiv(\phi_i)^*$.)
Aside from the terms
included in $L_{SB}$ in
Eq.~\Aaf, one might in general have $\ph^2\ph^*$-type couplings,
$\psi\psi$ mass terms or $\l\psi$-mixing terms (as long as
they satisfy a constraint that quadratic divergences are not produced).
However, the
soft-breaking terms we
have included are those which would be
engendered by an underlying supergravity theory
and which are therefore
considered most frequently in the literature.

The non-renormalisation theorem tells us that the
superpotential $W$ undergoes no infinite renormalisation
so that we have, for instance
\eqn\Ada{
\b_Y^{ijk}=Y^{ijp}\g^k{}_p+(k\leftrightarrow
i)+(k\leftrightarrow j),}
where $\g$ is the anomalous dimension for $\Ph$.
The one-loop results for the gauge coupling $\b$-function $\b_g$ and
for $\g$ are given by
\eqn\Aab{
\lf\b_g^{(1)}=g^3Q,\q\hbox{and}\q
\lf\g^{(1)i}{}_j=P^i{}_j,}
where
\eqna\Aac$$\eqalignno{
\lf Q&=T(R)-3C(G),\q\hbox{and}\q &\Aac a\cr
\lf P^i{}_j&={1\over2}Y^{ikl}Y_{jkl}-2g^2C(R)^i{}_j. &\Aac b\cr}$$
Here
\eqn\Aaca{
T(R)\d_{AB} = \Tr(R_A R_B),\q C(G)\d_{AB} = f_{ACD}f_{BCD} \q\hbox{and}\q
C(R)^i{}_j = (R_A R_A)^i{}_j.}
The one-loop $\b$-functions
for the soft-breaking couplings are given by
\eqna\Ac$$\eqalignno{
\lf\b_h^{(1)ijk}&=U^{ijk}+U^{kij}+U^{jki},&\Ac a\cr
\lf[\b_{m^2}^{(1)}]^i{}_j&=W^i{}_j+2g^2(R_A)^i{}_j\tr[R_Am^2],&\Ac b\cr
\lf\b_b^{(1)ij}&=V^{ij}+V^{ji},&\Ac c\cr
\lf\b_M^{(1)}&=2g^2QM,&\Ac d\cr}$$
where
\eqna\Aaa$$\eqalignno{
U^{ijk}&=h^{ijl}P^k{}_l+Y^{ijl}X^k{}_l,&\Aaa a\cr
V^{ij}&=b^{il}P^k{}_l+{1\over2}Y^{ijl}Y_{lmn}b^{mn}
+\m^{il}X^j{}_l,&\Aaa b\cr
W^j{}_i&={1\over2}Y_{ipq}Y^{pqn}(m^2)^j{}_n+{1\over2}Y^{jpq}Y_{pqn}(m^2)^n{}_i
+2Y_{ipq}Y^{jpr}(m^2)^q{}_r\cr &\quad
+h_{ipq}h^{jpq}-8g^2MM^* C(R)^j{}_i,&\Aaa c\cr}$$
with
\eqn\Aaba{
X^i{}_j=h^{ikl}Y_{jkl}+4Mg^2C(R)^i{}_j.}
Our assumption that the group $\cal G$ is semi-simple
implies that the $\tr[R_Am^2 ]$ term in Eq.~\Ac{b}\ is zero,
while the absence of gauge singlets means that
(for instance in Eq. \Aaa{b}) we have
\eqn\Aia{
Y_{ijk}b^{jk}=Y_{ijk}\m^{jk}=0. }
We then claim that the conditions
\eqna\Aj$$\eqalignno{h^{ijk}&=-MY^{ijk},&\Aj a\cr
(m^2)^i{}_j&={1\over3}(1-{1\over{\lf}}{2\over3}g^2Q)MM^*\d^i{}_j,&\Aj b\cr
b^{ij}&=-{2\over3}M\m^{ij}&\Aj c\cr}$$
are RG invariant through at least  two loops, provided we
impose the condition
\eqn\Ak{
P^i{}_j={1\over3}g^2Q\d^i{}_j.}
(The idea of seeking relations
amongst dimensionless couplings which are preserved by
renormalisation  has been explored  in the
coupling constant reduction programme of Zimmermann
{\it et al.}\ref\Zim{N.-P. Chang, \prd10 (1974) 2706\semi
J. Kubo, K. Sibold and W. Zimmermann, \npb259 (1985) 331\semi
R. Oehme, K. Sibold and W. Zimmermann, \plb153 (1985) 142\semi
K. Sibold and W. Zimmermann, \plb191 (1987) 427\semi
J. Kubo, K. Sibold and W. Zimmermann, \plb220 (1989) 185\semi
O. Piguet and K. Sibold, \plb229 (1989) 83\semi
W. Zimmermann, \plb311 (1993) 249.}.)
We first demonstrate the RG invariance of the conditions
Eq.~\Aj{}.
The invariance of Eq.~\Aj{a}\ requires
\eqn\An{
\b_h^{ijk}=-\b_MY^{ijk}-M\g^i{}_mY^{mjk}-M\g^j{}_mY^{imk}-M\g^k{}_mY^{ijm}. }
The strategy we adopt to verify equations such as Eq.~\An\ is to simplify
the $\b$-functions and
anomalous dimensions as follows: firstly we use Eq.~\Ak\ to replace $P^i{}_j$
by
$Q$. We also use Eqs. \Aj{}\ to replace $h^{ijk}$, $m^2$ and $b$ wherever they
occur. Having done
this, we find that any occurrences of $Y_{ikl}Y^{jkl}$, $C(R)$, $C(G)$ or
$T(R)$ can be written in terms of $P$
and $Q$ according to Eq.~\Aac{}. We can now use Eq.~\Ak\ again if necessary to
replace $P$ by $Q$.
For instance, we find, applying our strategy of imposing the condition
Eq.~\Aj{a}\ in Eq.~\Aaba, and using Eqs. \Aac{b}, \Ak,
\eqn\Ao{\eqalign{
X^i{}_j&=-MY^{ikl}Y_{jkl}+4g^2MC(R)^i{}_j\cr
       &=-{2\over3}g^2QM\d^i{}_j.\cr}}
Henceforth we shall simply assume that this procedure is
followed where possible.
For instance, from Eqs.~\Aaa{a}, \Ao, we find
\eqn\Ap{
U^{ijk}=-g^2QMY^{ijk} }
which, using Eqs.~\Ac{a,d}, ensures that Eq.~\An\ is satisfied at
one loop. The RG invariance of Eq.~\Aj{b}\ requires that
\eqn\Aq{
(\b_{m^2})^i{}_j={1\over3}([1-{1\over{\lf}}{2\over3}g^2Q][\b_MM^*+M(\b_M)^*]
-{1\over{\lf}}{4\over3}g\b_gQMM^*)\d^i{}_j. }
At one loop we readily find, from Eqs.~\Ac{b}, \Aaa{c},
\eqn\Ar{\eqalign{
W^i{}_j&=MM^*(4P^i{}_j-{1\over{\lf}}{2\over3}g^2QY_{ikl}Y^{jkl})\cr
&=g^2QMM^*({4\over3}\d^i{}_j-{1\over{\lf}}{2\over3}Y_{ikl}Y^{jkl}),}}
which, with Eqs.~\Ac{b,d}\ implies Eq.~\Aq\ at one loop. (The additional,
two-loop term in Eq.~\Ar\ will be required later.)
Finally, for the RG invariance of Eq.~\Aj{c}\ we need
\eqn\As{
\b_b^{ij}=-{2\over3}(\b_M\m^{ij}+M\g^i{}_k\m^{kj}+M\g^j{}_k\m^{ik}). }
 From Eqs.~\Aaa{b}, \Ao, we obtain
\eqn\At{
V^{ij}=-{8\over9}g^2QM\m^{ij},}
which, using Eqs.~\Ac{c,d}, leads immediately to Eq.~\As\ at one loop.
Finally, it behoves us to check that the condition Eq.~\Ak\ is itself
RG invariant. This amounts to the condition
\eqn\Al{
{1\over2}\left(\g^i{}_mY^{mkl}Y_{jkl}+\g^m{}_jY^{ikl}Y_{mkl}+4Y^{ikl}\g^m{}_l
Y_{jkm}\right)-4g\b_g C(R)^i{}_j={2\over3}g\b_gQ\d^i{}_j,}
which is easily verified at one loop using Eqs.~\Aab.
The fact that the conditions Eq.~\Aj{}, \Ak\ are preserved by renormalisation
at one loop seems to us remarkable enough; however, they are actually
preserved even at the two-loop level as well.
The two-loop
$\b$-functions for the dimensionless couplings were calculated in
Ref. \ref\two{D.R.T.~Jones, \npb87 (1975) 127\semi
A.J.~Parkes and P.C.~West, \plb138 (1984) 99;
             \npb256 (1985) 340\semi
             P. West, \plb137 (1984) 371\semi
          D.R.T. Jones and L. Mezincescu, \plb136 (1984) 242;
{\it ibid} 138 (1984) 293.}; they can be written in the form
\eqna\Au$$\eqalignno{ \llf\b_g^{(2)}&=2g^5C(G)Q-2g^3r^{-1}C(R)^i{}_jP^j{}_i
&\Au a\cr
\llf\g^{(2)i}{}_j&=[-Y_{jmn}Y^{mpi}-2g^2C(R)^p{}_j\d^i{}_n]P^n{}_p+
2g^4C(R)^i{}_jQ,&\Au b\cr}
$$
where $Q$ and $P^i{}_j$ are given by Eq.~\Aac{}, and $r=\d_{AA}$.

The calculation of the two-loop
$\b$-functions for the soft breaking couplings raises interesting
issues concerning the use of dimensional reduction in
non-\sic\ theories
\ref\jjr{I. Jack, D.R.T. Jones and K.L. Roberts,  \zpc 62 (1994) 161;
 {\it ibid} 63  (1994) 151.}.

The results are as follows\ref\mv{S.~P.~Martin
and M.~T.~Vaughn \plb 318 (1993) 331;
\prd50 (1994) 2282.}\nref\yamada{Y.~Yamada, \plb 316 (1993) 109;
\prl72 (1994) 25; \prd50 (1994) 3537.}\nref\jj{I.~Jack and
D.R.T.~Jones, \plb 333 (1994) 372.}--\ref\jjmvy{I.~Jack,
D.R.T.~Jones,
    S.P.~Martin, M.T.~Vaughn  and Y.~Yamada, \prd 50 (1994) R5481.}:
\eqna\Av$$\eqalignno{
\llf\b_h^{(2)ijk}&=-\Bigl[h^{ijl}Y_{lmn}Y^{mpk}+2Y^{ijl}Y_{lmn}
h^{mpk}-4g^2MY^{ijp}C(R)^k{}_n\Bigr]P^n{}_p\cr&\quad-2g^2U^{ijl}C(R)^k{}_l
+g^4(2h^{ijl}-8MY^{ijl})C(R)^k{}_lQ-Y^{ijl}Y_{lmn}Y^{pmk}X^n{}_p\cr
&\quad+(k\leftrightarrow i)+(k\leftrightarrow j),&\Av a\cr
\llf[\b_{m^2}^{(2)}]^j{}_i
&=\biggl(-\Bigl[(m^2)_i{}^lY_{lmn}Y^{mpj}
+{1\over2}Y_{ilm}Y^{jpm}(m^2)^l{}_n+{1\over2}Y_{inm}Y^{jlm}(m^2)^p{}_l
\cr&\quad+Y_{iln}Y^{jrp}(m^2)^l{}_r+h_{iln}h^{jlp}\cr&\quad+
4g^2MM^*C(R)^j{}_n\d^p{}_i+2g^2(R_A)^j{}_i(R_Am^2)^p{}_n\Bigr]P^n{}_p\cr
&\quad+\bigl[2g^2M^*C(R)^p{}_i\d^j{}_n-h_{iln}Y^{jlp}\bigr]X^n{}_p
-{1\over2}\bigl[Y_{iln}Y^{jlp}+2g^2C(R)^p{}_i\d^j{}_n\bigr]W^n{}_p\cr
&\quad +12g^4MM^* C(R)^j{}_iQ+4g^4SC(R)^j{}_i\biggr)+{\rm h.c.},&\Av b\cr
\llf\b_b^{(2)ij}&=\Bigl[-b^{il}Y_{lmn}Y^{mpj}
-2\m^{il}Y_{lmn}h^{mpj}-Y^{ijl}Y_{lmn}b^{mp}
\cr&\quad+4g^2MC(R)^i{}_k\m^{kp}\d^j{}_n\Bigr]P^n_p
-\bigl[\m^{il}Y_{lmn}Y^{mpj}+{1\over2}Y^{ijl}Y_{lmn}\m^{mp}\bigr]X^n{}_p
\cr&\quad-2g^2C(R)^i{}_kV^{kj}+g^2C(R)^i{}_kY^{kjl}Y_{lmn}b^{mn}\cr &\quad
+2g^4(b^{ik}-4M\m^{ik})C(R)^j{}_kQ+(i\leftrightarrow j),&\Av c\cr
\llf\b_M^{(2)}&=g^2\Bigl(8g^2C(G)QM-4r^{-1}C(R)^i{}_jP^j{}_iM
+2r^{-1}X^i{}_jC(R)^j{}_i\Bigr),&\Av d\cr}$$
where
\eqn\Ava{
S\d_{AB}=(m^2)^i{}_j(R_AR_B)^i{}_j-MM^*C(G)\d_{AB}.}
The expressions given in Eq.~\Av{} (and in particular Eq.~\Av{b}) correspond
to the use of a particular subtraction scheme whereby the mass of the
$\e$-scalars decouples from the evolution of the other parameters. For
a discussion, see refs.~\jj , \jjmvy .

At two loops we find, applying the usual procedure to Eqs.~\Au{}, \Av{d},
\eqna\Aw$$\eqalignno{
\llf\g^{(2)i}{}_j&=-{2\over9}g^4Q^2\d^i{}_j,&\Aw a\cr
\llf\b^{(2)}_g&=-{2\over3}g^5Q^2,&\Aw b\cr
\llf\b^{(2)}_M&=-{8\over3}g^4Q^2M.&\Aw c}$$
Now we can
go on to check the RG invariance of Eqs.~\Aj{}\ to two-loop order. Using
Eqs.~\Ao, \Ap\ in Eq.~\Av{a}, we find
\eqn\Ax{
\llf\b^{(2)ijk}_h={10\over3}g^4Q^2MY^{ijk}.}
Inserting Eqs.~\Aw{a,c}, and \Ax\ into Eq.~\An, we immediately
verify the two-loop RG-invariance of Eq.~\Aj{a}.
Now using Eqs. \Ao, \Ar\ in Eq.~\Av{b}, we obtain
\eqn\Az{
\llf(\b_{m^2}^{(2)})^i{}_j=4Qg^2MM^*(-Y_{ikl}Y^{jkl}+{14\over3}g^2C(R)^i{}_j).}
Hence, from Eqs.~\Ac{b}, \Ar, \Az, we obtain
\eqn\Aza{
(\b_{m^2}^{(1)}+\b_{m^2}^{(2)})^i{}_j={1\over{\lf}}Qg^2MM^*({4\over3}-
{1\over{\lf}}{28\over9}g^2Q).}
Using Eqs.~\Aia, \Aza, \Aw{c}, \Ac{d}, \Aab\ in Eq.~\Aq, we see that
Eq.~\Aj{b}\ is RG invariant  throughtwo loops.
Using Eqs.~\Ao, \At, \Ak, \Aab\ in Eq.~\Av{c},
we find
\eqn\Azb{
\llf\b^{(2)ij}_b={56\over27}g^4Q^2M\m^{ij}. }
On substituting Eqs.~\Azb\ and \Aw{a,c}\ into Eq.~\As, we see that
Eq.~\Aj{c}\ is RG invariant at two loops.
Finally, using Eqs. \Aw{a,b}, we verify Eq.~\Al\ at
two loops, ensuring the RG invariance of Eq.~\Ak\ at this level.
Thus we have demonstrated the RG invariance of Eqs. \Aj{}\ and \Ak\ through
two loops.

We turn now to the possibility of constructing realistic models
satisfying our constraints. The main impact on  low-energy physics, is
that from Eq.~\Aj{}\ we have (in the usual notation)
a universal scalar mass $m_0$  and universal $A$ and $B$ parameters
related (to lowest order in $g^2$) to the gaugino mass $M$ as follows:
\eqna\Baa$$\eqalignno{
 m_0 &= {1\over{\sqrt{3}}}M,&\Baa a\cr
A &= -M,&\Baa b\cr
B &= -{2\over3}M.&\Baa c\cr}$$
Evidently it will be interesting to explore the region of
the usual \ssm\ parameter space consistent with Eq.~\Baa{};
current experimental constraints will probably
not rule out the scenario {\it per se\/}, but the various
super-partner masses will be more tightly correlated
than in the usual approach.

It follows from our results that if $P^i{}_j=Q=0$, (guaranteeing that the
dimensionless coupling $\b$-functions are zero to two loops) then
soft-breaking couplings related by Eq.~\Baa{}\ will also have vanishing
$\b$-functions, leading to the possibility of
finite softly-broken \sic\ theories. This has already been pointed out at the
one-loop level in Ref.~\ref\jmy{D.R.T.~Jones,
L.~Mezincescu and Y.-P.~Yao, \plb148 (1984) 317.}\ and at
the two-loop level in Ref. \jj. In Ref.~\jmy\
it was remarked that Eqs.~\Baa{a,b}\ are
consistent with the pattern of soft-breaking terms which
emerges from \sy\ breaking in the hidden sector of an underlying
supergravity theory with a ``minimal'' K\"ahler potential.
Even more interestingly, Eqs.~\Baa{a,b}\
are identical to relations which arise in effective supergravity
theories motivated by superstring theory, where \sy\ breaking is assumed to
occur purely via a vacuum expectation value for the dilaton\IL\CM.
More general scenarios involving
vacuum expectation values for other moduli fields are also possible. To be more
specific, we follow
Ref.~\CM\ in concentrating on the modulus $T$ whose classical value gives the
size of the manifold, and parametrising the ratio of the auxiliary fields
$F^S$ and $F^T$ for the dilaton $S$ and for $T$ by an angle $\th$--so that
$\th$ characterises the extent to which supersymmetry-breaking is dominated
by $S$ or $T$.
We also simplify still
further by assuming a vanishing cosmological constant and by ignoring string
loop corrections, and also the phases
of $F^S$ and $F^T$. In this more general case, the gaugino mass is related to
the gravitino mass $m_{3\over2}$ by $M=\sqrt3m_{3\over2}\sin\th$, and
the soft-breaking parameters $m_0$ and $A$ are still given by
Eqs.~\Baa{a,b}, while
$B$ is either given by\CM
\eqn\Bab{
B=-{M(1+\sqrt3\sin\th+\cos\th)\over{\sqrt3\sin\th}}}
or\ref\err{A. Brignole, L. E. Ib\'a\~nez and C. Mu\~noz,
erratum to \npb422 (1994) 125.}
\eqn\Bac{
B={2M\over{\sqrt3\sin\th}},}
depending on whether the $\m$ term is generated by an explicit $\m$-term in
the supergravity superpotential, or by a special term in the K\"ahler
potential. In the first case the value $\th={4\pi\over3}$
reproduces our constraint Eq.~\Baa{c}; however, in the second case there is
no value of $\th$ which is consistent with this constraint.

In addition to Eq. \Baa{}, we also need to
to impose Eq.~\Ak\ as a condition on the theory at the unification scale.
It is not clear at present how such a constraint would naturally emerge from
string theory.
The special case $P^i{}_j=Q=0$, corresponding, as we have remarked, to
two-loop finite theories, was tabulated in
Ref.~\ref\HPS{S.~Hamidi, J.~Patera and J.H.~Schwarz, \plb 141 (1984) 349.}.
They found a fair number of possibilities, including a few
of phenomenological interest: in particular
a simple $SU_5$ model
\ref\jr{D.R.T.~Jones and S.~Raby, \plb 143 (1984) 137.}
\ref\utc{S.~Hamidi and J.H.~Schwarz, \plb 147 (1984) 301\semi
J.E.~Bj\"orkman,  D.R.T.~Jones and S. Raby, \npb 259 (1985) 503\semi
J. Leon et al,  \plb 156 (1985) 66\semi
D.R.T.~Jones and A.J.~Parkes, \plb160 (1985) 267\semi
D.R.T.~Jones, \npb 277 (1986) 153\semi
A.V.~Ermushev,   D.I.~Kazakov and  O.V.~Tarasov,
\npb 281 (1987) 72\semi
D.~Kapetanakis, M.~Mondragon and G.~Zoupanos, \zpc 60 (1993) 181\semi
M.~Mondragon and G.~Zoupanos, CERN-TH.7098/93\semi
N.G.~Deshpande, Xiao-Gang~He and E.~Keith, \plb 332 (1994) 88\semi
J.~Kubo,  M.~Mondragon and G.~Zoupanos,
\npb 424 (1994) 291.}.

We can anticipate, therefore,
that it will be possible to construct unified
models satisfying Eq.~\Ak . The obvious try is a simple generalisation of the
finite $SU_5$ model first analysed in ref.~\jr .  The superpotential is
\eqn\jrw{
W = {1\over2}A_{ij\a} 10_i 10_j H_{\a} + B_{ij\a}10_i\5bar_j\Hbar_{\alpha}
    +C_{\a\b}\Hbar_{\a}24 H_{\b} + D 24^3}
where $i,j: 1\ldots x$ and $\a,\b : 1\ldots y$ so that we have
$x$ generations, $y$ sets of Higgs multiplets $(H + \Hbar)$ and a single
adjoint $(24)$. It is straightforward to write down Eq.~\Ak\ for this model;
tracing on all indices we obtain the relations:

\eqna\jrmod$$\eqalignno{
|A|^2 + {8\over5}|C|^2 &= g^2 y\left({8\over5}+{1\over9}Q\right),&\jrmod   a\cr
|B|^2 + {6\over5}|C|^2 &= g^2 y\left({6\over5}+{1\over{12}} Q\right),
&\jrmod   b\cr
|B|^2 &= g^2 x\left({6\over5}+{1\over{12}} Q\right),&\jrmod   c\cr
3|A|^2 + 2|B|^2 &= g^2 x\left({36\over5}+{1\over{3}} Q\right),&\jrmod   d\cr
{189\over5}D^2 + C^2 &= g^2 \left( 10 +{1\over{3}} Q\right).&\jrmod   e\cr}$$
where here $Q = 2x + y - 10$. It is easy to show, however, that Eqs.~\jrmod{}
do not have a solution unless $Q=0$, which corresponds to the finite
case. This outcome is not generic, however, and it is easy to modify
the theory so as to produce candidate theories that do satisfy Eq.~\Ak ,
for example by including one or more sets of $10 + \10bar$
multiplets.  It remains to be seen, however, whether there
exists a compelling unified theory with universal soft breaking terms.
Meanwhile, if we conjecture that such a theory leads to the same
low energy physics as the \ssm , we can at least explore the consequences
of Eq.~\Baa{} for the super-particle spectrum. We will report on these
calculations elsewhere.

\vskip 20pt
\line{\bf Acknowledgements\hfil}
We thank Luis Ib\'a\~nez for communications, and in particular for drawing
Ref~\err\ to our attention.
IJ thanks Carlos Mu\~noz and Dennis Silverman for useful
conversations, and also thanks PPARC for financial support.

\listrefs
\bye